\documentclass[preprint,showpacs,preprintnumbers,amsmath,amssymb,amssymb]{revtex4}
\usepackage{graphicx}
\usepackage{dcolumn}
\usepackage{bm}
\begin{document}
\title{Probing doubly charged Higgs in $e^+ e^-$ Colliders in 3-3-1 Model}  
\author{J. E. Cieza Montalvo$^1$, Nelson V. Cortez$^2$, M. D. Tonasse$^3$}   
\address{$^1$Instituto de F\'{\i}sica, Universidade do Estado do Rio de Janeiro, Rua S\~ao Francisco Xavier 524, 20559-900 Rio de Janeiro, RJ, Brazil}
\address{$^2$Justino Boschetti 40, 02205-050 S\~ao Paulo, SP, Brazil}
\address{$^3$Unidade de Registro, {\it Campus} Experimental de Registro, Universidade Estadual Paulista, Rua Tamekishi Takano 5, 11900-000, Registro, SP, Brazil}  
\date{\today}
\begin{abstract}
The SU(3)$_L\otimes$U(1)$_N$ electroweak model predicts new Higgs bosons beyond the one of the standard model. In this work we investigate the signature and production of doubly charged Higgs bosons in the $e^-e^+$ 
International Linear Collider and in the CERN Linear Collider.  
We compute the branching ratios for the doubly charged gauge bosons of the model.

\end{abstract}
\maketitle
\section{INTRODUCTION \label{introd}}
A way to the understanding of the symmetry breaking in particle physics is through of the scalars. These particles protect the unitarity of the theory, in the standard model(SM) and other models, by moderating the cross section growth and generating masses to all particles. The discovery of these scalars will be a crucial experiment of the SM and beyond it. Some extensions of the SM have two Higgs doublet, which arises naturally in models as supersymmetric one, but also there is (are) Higgs triplet(s) as in models like $SU(3)_C \otimes SU(3)_L \otimes U(1)_N$ (3-3-1)\cite{PP92,PT93a}, Left-right symmetric (LRSM)\cite{pati,rizzo82}, Higgs triplet \cite{georgi}, Little Higgs \cite{arkani} etc. these models with Higgs trplet predict doubly charged Higgs bosons (DCHBs) and in several of them, these new particles appear as being relatively light. However DCHBs there exist also as components of Higgs sextet, as for instance in \cite{Foot93}.\par
The Yukawa couplings are usually the manner where Higgs triplet(s) is (are) coupled to matter field and generally this kind of triplet have no relation with the fermion masses, although that in the model used here it is responsible to generate mass to the exotic lepton. \par
In some models the DCHBs do not couple to quarks and their coupling to lepton break the lepton number by two units. As a result, these new scalar particles have a distinct experimental signature, namely a same-sign lepton pairs. In models as supersymmetric left-right models, two triplet Higgs fields are used to break the SU(2) gauge symmetry, in this case, the double triplet, is called left- and right-handed. \par
In addition, the presence of DCHBs provides a simple explanation to the lightness of neutrinos masses through seesaw mechanism \cite{mohapa} according to recent data on neutrino oscillation \cite{ahn03}. For all reasons above, models with Higgs triplet is very important for nowadays {\it state of art} in the searching for a more complete universal particle model. \par\par
Important searches have been proposed, as for instance, in hadron colliders \cite{huitu},  in LEP at CERN,  as OPAL detector, including Bhabha scattering \cite{abbiendi02} or in L3 and Delphi groups \cite{achard03}.
In HERA at DESY,  through $ep$ collision \cite{H106} and several others collision experiments as $\gamma\gamma$ collision \cite{surajit98}, $e\gamma$ collision \cite{Godfrey02} as in SLAC, or even {\it via} $Z \to H^{++} H^{--}$ \cite{swartz90} or even through $W$ fusion \cite{vega90}.\par
Since the beginning of eighties decade, the usual model used for the DCHB search has been the LRSM. It has been argued that such a DCHBs could occour in that model naturally associated with majorana neutrinos, where the discovery of these kind of particle would be a signal of flavor lepton violation by two units in several processes \cite{rizzo82}. after almost forty 
years of researches news models have been proposed with the DCHBs inside them, which deserved be tested. Other model extending the gauge sector of 3-3-1 models has been proposed \cite{coton05} and analysis of scalar sector is being investigated using inclusive exact solution proposed recently \cite{apalcu06}. \par
The main motivation of our work is to show that in the context of 3-3-1 model the signatures for doubly charged Higgs $H^{\pm \pm}$ is very significant for the International Linear collider (ILC) and CERN LInear Collider (CLIC). Our continuos results have indicated a satisfactory number of events to establish the signal in several kinds of particle colliders and through of the analyze of those results we can infer about the existence of DCHB and heavy leptons \cite{cnt1,cnt2,cnt3}.
\section{Overview of the Model \label{sec2}}
The underlying electroweak symmetry group is SU(3)$_L$$\otimes$U(1)$_N$, where $N$ is the quantum number of the U(1) group. Therefore, the left-handed lepton matter content is $\left(\begin{array}{ccc} \nu^\prime_a & \ell^\prime_a & 
{\tt L}^\prime_a\end{array}\right)^{\tt T}_L$ transforming as $\left({\bf 3},
0\right)$, where $a = e, \mu, \tau$ is a family index (we are using
primes for the interaction eigenstates). ${\tt L}^\prime_{aL}$ are lepton 
fields which can be the charge conjugates ${\ell^\prime_{aR}}^C$
\cite{PP92}, ${\nu^\prime_{La}}^C$ are the antineutrinos or heavy leptons $P^{\prime+}_{aL}$ $\left(P^{\prime+}_{aL} =
E^{\prime+}_{L}, M^{\prime+}_{L}, T^{\prime+}_{L}\right)$ 
\cite{PT93a}.  \par 
The model of Ref. \cite{PT93a} has the simplest scalar sector for 3-3-1 Models. In this version the charge operator is given by 
\begin{equation}
\frac{Q}{e} = \frac{1}{2} \left(\lambda_3 - \sqrt{3}\lambda_8 \right)+ N , 
\label{carga}
\end{equation}\noindent
where $\lambda_3$ and $\lambda_8$ are the diagonal Gell-Mann matrices and $e$ is the elementary electric charge. The right-handed charged leptons are introduced in singlet
representation of SU(3)$_L$ as $\ell^{\prime -}_{aR} \sim \left({\bf
1}, -1\right)$ and $P^{\prime +}_{aR} \sim \left({\bf 1}, 1\right)$.\par
The quark sector is given by
\begin{equation}
Q_{1L} = \left(\begin{array}{c} u^\prime_1 \cr d^\prime_1 \cr J_1 
\end{array}\right)_L \sim \left({\bf 3}, \frac{2}{3}\right), \qquad
Q_{\alpha L} = \left(\begin{array}{c} d^\prime_\alpha \cr
u^\prime_\alpha \cr J^\prime_\alpha \end{array}\right)_L \sim
\left({\bf 3}^*, -\frac{1}{3}\right), 
\end{equation}
where $\alpha = 2, 3$, $J_1$ and $J_\alpha$ are exotic quarks with electric charge  $5/3$ and $4/3$ respectively. It must be notice that the first quark family transforms differently from the two others under the gauge group, which is essential for the anomaly cancellation mechanism \cite{PP92}. \par 
The physical fermionic eigenstates rise by the transformations
\begin{subequations}\begin{eqnarray}
&& \ell^{\prime -}_{aL,R} = A^{L,R}_{ab}\ell^-_{bL,R}, \quad
P^{\prime +}_{aL,R} = B^{L,R}_{ab}P^+_{bL,R}, \\ && U^\prime_{L, R} 
= {\cal U}^{L, R}U_{L, R}, \quad D^\prime_{L, R} = {\cal D}^{L,
R}D_{L, R}, \quad J^\prime_{L, R} = {\cal J}^{L, R}J_{L, R},
\end{eqnarray}\label{eigl}\end{subequations}
where $U_{L, R} = \left(\begin{array}{ccc} u & c & t 
\end{array}\right)_{L, R}$, $D_{L, R} = \left(\begin{array}{ccc} d &
s & b \end{array}\right)_{L, R}$, $J_{L, R} =
\left(\begin{array}{ccc} J_1 & J_2 & J_3 
\end{array}\right)_{L, R}$ and $A^{L, R}$, $B^{L, R}$, ${\cal U}^{L, R}$, ${\cal D}^{L, R}$, ${\cal J}^{L, R}$ are arbitrary mixing matrices. \par 
The minimal scalar
sector contains the three scalar triplets
\begin{equation}
\eta = \left(\begin{array}{c} \eta^0 \\ \eta_1^- \\ \eta_2^+
\end{array}\right) \sim \left({\bf 3}, 0\right), \quad \rho =
\left(\begin{array}{c} \rho^+ \\ \rho^0 \\ \rho^{++} 
\end{array}\right) \sim \left({\bf 3}, 1\right), \quad \chi =
\left(\begin{array}{c} \chi^- \\
\chi^{--} \\ \chi^0 \end{array}\right) \sim \left({\bf 3}, -1\right).
\label{eigh}
\end{equation}
The most general, gauge invariant and renormalizable Higgs potential, which conserves the leptobaryon number \cite{PT93b}, is 
\begin{eqnarray}
V\left(\eta, \rho, \chi\right) & = & \mu_1^2\eta^\dagger\eta +
\mu_2^2\rho^\dagger\rho + \mu_3^2\chi^\dagger\chi +
\lambda_1\left(\eta^\dagger\eta\right)^2 +
\lambda_2\left(\rho^\dagger\rho\right)^2 + 
\lambda_3\left(\chi^\dagger\chi\right)^2 + \cr && +
\left(\eta^\dagger\eta\right)
\left[\lambda_4\left(\rho^\dagger\rho\right) +
\lambda_5\left(\chi^\dagger\chi\right)\right] + \lambda_6
\left(\rho^\dagger\rho\right)\left(\chi^\dagger\chi\right) + 
\lambda_7\left(\rho^\dagger\eta\right)\left(\eta^\dagger\rho\right)
+ \cr && +
\lambda_8\left(\chi^\dagger\eta\right)\left(\eta^\dagger\chi\right)
+
\lambda_9\left(\rho^\dagger\chi\right)\left(\chi^\dagger\rho\right) 
+ \frac{1}{2}\left(f\epsilon^{ijk}\eta_i\rho_j\chi_k + {\mbox{H.
c.}}\right) \label{pot}\end{eqnarray} 
The neutral components of the scalars triplets (\ref{eigh}) develop non zero vacuum expectation value (VEV) $\langle\eta^0\rangle = v_\eta$, $\langle\rho^0\rangle = v_\rho$ and 
$\langle\chi^0\rangle = v_\chi$, with $v_\eta^2 + v_\rho^2 = v_W^2 =
(246 \mbox{ GeV})^2$. This mechanism  generate the fermion and gauge boson masses \cite{ton1}. The pattern of symmetry breaking is $\mbox{SU(3)}_L 
\otimes\mbox{U(1)}_N\stackrel{\langle\chi\rangle}{\longmapsto}
\mbox{SU(2)}_L\otimes\mbox{U(1)}_Y\stackrel{\langle\eta,
\rho\rangle}{\longmapsto}\mbox{U(1)}_{\rm em}$. Therefore, we can
expect $v_\chi \gg v_\eta, v_\rho$. In the potential (\ref{pot}), $f$ 
and $\mu_j$ $\left(j = 1, 2, 3\right)$ are  constants with dimension of mass and the $\lambda_i$ $\left(i =
1, \dots, 9\right)$ are adimensional constants with $\lambda_3 < 0$
and $f < 0$ from the positivity of the scalar masses \cite{PT93b}. The $\eta$ and $\rho$ scalar triplets give masses to the ordinary fermions and gauge bosons, while the $\chi$ scalar triplet gives masses to the new fermion and gauge bosons. In this work we are using the eigenstates and masses of the Ref. \cite{PT93b}. For others analysis of the 3-3-1 Higgs potential see Ref. \cite{DL06}. \par 
Symmetry breaking is initiated when the scalar
neutral fields are shifted as $\varphi = v_\varphi + \xi_\varphi +
i\zeta_\varphi$, with $\varphi$ $=$ $\eta^0$, $\rho^0$, $\chi^0$.
Thus, the physical neutral scalar eigenstates $H^0_1$, $H^0_2$, 
$H^0_3$ and $h^0$ are related to the shifted fields as
\begin{subequations}\begin{equation}
\left(\begin{array}{c} \xi_\eta \\ \xi_\rho \end{array}\right)
\approx \frac{1}{v_W}\left(\begin{array}{cc} c_\omega & s_\omega \\ 
s_\omega & -c_\omega
\end{array}\right)\left(\begin{array}{c} H^0_1 \\
H^0_2 \end{array}\right), \qquad \xi_\chi \approx H^0_3, \qquad
\zeta_\chi \approx ih^0, \label{eign}\end{equation} 
and in the charged scalar sector we have 
\begin{eqnarray}
&& \eta^+_1 = s_\omega H^+_1, \qquad \eta^+_2 = s_\varphi H_2^+,
\qquad \rho^+ = c_\omega H_1^+, \\ && \chi^+ = c_\varphi H^+_2,
\qquad \rho^{++} = s_\phi H^{++}, \qquad \chi^{++} = c_\phi H^{++}, 
\label{eigc}\end{eqnarray}\label{eig}\end{subequations} 
with the condition that $v_\chi \gg v_\eta$, $v_\rho$ in Eqs. (\ref{eign})
and $c_\omega$ = $\cos\omega$ = $v_\eta$ / $\sqrt{v_\eta^2 + v_\rho^2}$,
$s_\omega = \sin\omega$, $c_\phi = \cos\phi = v_\rho / \sqrt{v_\rho^2 + v_\chi^2}$, $s_\phi = \sin\phi$, $c_\varphi = \cos\varphi$ = $v_\eta / \sqrt{v_\eta^2 + v_\chi^2}$, $s_\varphi = \sin\varphi$. The $H_1^0$ Higgs boson in Eq. (\ref{eign}) looks like the standard model scalar boson, since its mass has no dependence on the VEV $v_\chi$ \cite{PT93b}. \par
The Yukawa interactions for leptons and quarks are, respectively,
\begin{subequations}\begin{eqnarray}
-{\cal L}_\ell & = & G_{ab}\overline{\psi}_{aL}\ell^\prime_{bR}\rho
+ G^\prime_{ab}\overline{\psi}_{aL}P^\prime_{bR}\chi + {\mbox{H. 
c.}}, \label{Ll} \\ 
-{\cal L}_Q & = &
\overline{Q}_{1L}\sum_i\left[G^u_{1i}U^\prime_{iR}\eta +
G^d_{1i}D^\prime_{iR}\rho + \sum_\alpha\overline{Q}_{\alpha
L}\left(F^u_{\alpha i}U^\prime_{iR}\rho^* + F^d_{\alpha 
i}D^\prime_{iR}\eta^*\right)\right] + \cr && +
G^j\overline{Q}_{1L}J_{1R}\chi +
\sum_{\alpha\beta}G^j_{\alpha\beta}\overline{Q}_{\alpha
L}J^\prime_{\beta R}\chi^* + {\mbox{H. c.}}
\end{eqnarray}\label{Lq}\end{subequations} 
In Eqs. (\ref{Lq}), as before mentioned $a,b = e, \mu, \tau$  and $\alpha = 2,3$. \par 
Beyond the standard particles $\gamma$, $Z$ and $W^\pm$ the model predicts, in the gauge sector, one neutral $\left(Z^\prime\right)$, two single charged $\left(V^\pm\right)$ and two double charged $\left(U^{\pm\pm}\right)$ gauge bosons. The interactions between the gauge and Higgs bosons are given by the covariant derivative 
\begin{equation}
{\cal D}_\mu\varphi_i = \partial_\mu\varphi_i -
ig\left(\vec{W}_\mu.\frac{\vec{\lambda}}{2}\right)^j_i\varphi_j -
ig^\prime N_\varphi\varphi_iB_\mu,
\label{cov}\end{equation}
where $N_\varphi$ are the U(1) charges for the $\varphi$ Higgs 
triplets $\left(\varphi = \eta, \rho, \chi\right)$. $\vec{W}_\mu$
and $B_\mu$ are field tensors of SU(2) and U(1), respectively,
${\vec{\lambda}}$ are Gell-Mann matrices and $g$ and $g^\prime$ are
coupling constants for SU(2) and U(1), respectively. \par 
Introducing the eigenstates (\ref{eigl}) and (\ref{eig}) in the Lagrangeans (\ref{Lq}) we obtain the Yukawa interactions as function of the physical eigenstates, {\it i. e.},
\begin{subequations}\begin{eqnarray}
-{\cal L}_\ell & = &
\frac{1}{2}\left\{\frac{1}{v_\rho}\left[c_\omega\overline{\nu}{\cal
U}^{\nu e}H^+_1 + \left(v_\rho + s_\omega H_1^0 - c_\omega
H_2^0\right)\overline{e^-} + s_\phi \overline{P^+}{\cal
U}^{Pe}H^{++}\right]M^eG_Re^- + \right. \cr && \left. +
\frac{1}{v_\chi}\left[c_\omega\overline{\nu}{\cal V}^{\nu P}H_2^- +
c_\phi\overline{e^-}{\cal V}^{eP}H^{--} + \left(v_\chi + H_3^0 +
ih^0\right)\overline{P^+}\right]M^EG_RP^+\right\} + \cr && + 
{\mbox{H. c.}}, \label{llep}\\ 
-{\cal L}_Q & = &
\frac{1}{2}\left\{\overline{U}G_R\left[1 + \left[\frac{s_\omega}{v_\rho}
+ \left(\frac{c_\omega}{v_\eta} +
\frac{s_\omega}{v_\rho}\right){\cal V}^u\right]H_1^0 + \left[- 
\frac{c_\omega}{v_\rho} + \left(\frac{s_\omega}{v_\eta} -
\frac{c_\omega}{v_\rho}\right){\cal V}^u\right]H_2^0\right]M^uU +
\right. \cr && \left. + \overline{D}G_R\left[1 +
\left[\frac{c_\omega}{v_\eta} + \left(\frac{s_\omega}{v_\rho} - 
\frac{c_\omega}{v_\eta}\right){\cal V}^D\right]H_1^0 +
\left[\frac{s_\omega}{v_\eta} - \left(\frac{c_\omega}{v_\rho} +
\frac{s_\omega}{v_\eta}\right){\cal V}^D\right]H_2^0\right]M^dD +
\right. \cr && + \left. \overline{U}G_R\left[\frac{s_\omega}{v_\eta}V^\dagger_{\rm CKM}H^-_1 
+ \left(\frac{c_\omega}{v_\eta} -
\frac{s_\omega}{v_\rho}\right){\cal V}^{ud}H^+_1\right]M^dD +
\right. \cr && + \left.
\overline{D}G_R\left[\frac{c_\omega}{v_\rho}V_{\rm CKM}H^+_1 +
\left(\frac{s_\omega}{v_\eta} - \frac{c_\omega}{v_\rho}\right){\cal 
V}^{ud\dagger}H_1^-\right]M^uU\right\} + {\mbox{H. c.}}, \label{lqua}\\ 
-{\cal
L}_J & = & \frac{1}{2}\left[\overline{J}G_R{\cal
J}^{L\dagger}\left({\cal NU}^LM^uU + {\cal RD}^LM^dD\right) +
\right. \cr && \left. + \left(\overline{U}{\cal U}^{L\dagger}{\cal 
X}_1 + \overline{D}{\cal D}^{L\dagger}{\cal X}_2 + \overline{J}{\cal
J}^{L\dagger}{\cal X}_0\right){\cal J}^LM^JG_RJ\right] + {\mbox{H.
c.}}, \label{yukj}
\end{eqnarray}\label{yukt}\end{subequations}
where $G_R = 1 + \gamma_5$, $V_L^UV_L^D = V_{\rm CKM}$, the 
Cabibbo-Kobayashi-Maskawa mixing matrix, ${\cal U}^{\nu e}$, ${\cal
U}^{Pe}$, ${\cal V}^{\nu e}$, ${\cal V}^{eP}$, ${\cal V}^u =
V_L^U\Delta V_L^{U\dagger}$, ${\cal V}^d = V_L^D\Delta
V_L^{D\dagger}$ and ${\cal V}^{ud} = V_L^U\Delta V_L^{D\dagger}$ are arbitrary mixing 
matrices, $M^e = {\mbox{diag}}\left(\begin{array}{ccc} m_e & m_\mu &
m_\tau \end{array}\right)$, $M^P =
{\mbox{diag}}\left(\begin{array}{ccc} m_E & m_M & m_T
\end{array}\right)$, $M^u = {\mbox{diag}}\left(\begin{array}{ccc} 
m_u & m_c & m_t \end{array}\right)$, $M^d =
{\mbox{diag}}\left(\begin{array}{ccc} m_d & m_s & m_b
\end{array}\right)$ and $M^J = {\mbox{diag}}\left(\begin{array}{ccc}
m_{J_1} & m_{J_2} & m_{J_3} \end{array}\right)$. In Eq. (\ref{yukj}) 
we have defined
\begin{subequations}\begin{eqnarray}
{\cal N} = \left(\begin{array}{ccc} s_\omega H_2^+/v_\eta & 0
&             0        \cr
                              0            & s_\phi H^{--}/v_\rho & s_\phi H^{--}/v_\rho \cr 
                              0            & s_\phi H^{--}/v_\rho & s_\phi H^{--}/v_\rho \end{array}\right), \quad  {\cal X}_0 \approx \frac{v_\chi + H_3^0 + ih^0}{v_\chi}\left(\begin{array}{ccc} 1 & 0 & 0 \cr 
                                                          0 & 1 & 1 \cr
                                                          0 & 1 & 1 \end{array}\right), &&\\
{\cal R} = \left(\begin{array}{ccc}   s_\phi H^{++}/v_\rho   &           0       &        0          \cr 
                 0         & s_\omega H_2^-/v_\eta & s_\omega H_2^-/v_\eta \cr
                 0         & s_\omega H_2^-/v_\eta & s_\omega H_2^-/v_\eta \end{array}\right), \quad {\cal X}_1 = \frac{1}{v_\chi}\left(\begin{array}{ccc} c_\omega H_2^- &        0      &      0       \cr 
              0          & c_\phi H^{++} & c_\phi H^{++} \cr
              0          & c_\phi H^{++} & c_\phi H^{++} \end{array}\right), &&\\
{\cal X}_2 = \frac{1}{v_\chi}\left(\begin{array}{ccc} c_\phi H^{--} &         0       &        0        \cr 
              0        & c_\omega H_2^+ & c_\omega H_2^+ \cr
              0        & c_\omega H_2^+ & c_\omega H_2^+ \end{array}\right).&&
\end{eqnarray}\label{lqj}
\end{subequations} 
We call attention to the fact that non standard field interactions violate leptonic number, as can be seen from the Lagrangians (\ref{Lq}) and (\ref{cov}). However the total leptonic number is conserved \cite{PP92}.
\section{CROSS SECTION PRODUCTION}
\label{secIV}
\subsection{$e^{-} e^{+} \rightarrow H^{\pm \pm} H^{\mp \mp}$}
The production of DCHBs particles in $e^+e^-$ collisions occurs in association with the boson $\gamma$, $Z$,  $Z^{'}$, $H_{1}^{0}$ and $H_{2}^{0}$ in the s channel. This production mechanism can be studied 
through the analysis of the reactions $e^{-} e^{+} \rightarrow H^{\pm \pm} H^{\mp \mp}$, provided if there is enough available energy ($\sqrt{s} \geq 2m_{\pm \pm}$, where $m_{\pm \pm}$ is the mass of DCHB ). There is anrother contribution coming from t-channel, that is $e^{-}e^{+} \rightarrow H^{++} H^{--}$ via t-channel heavy lepton exchange, this contribution are three to four orders of magnitude smaller than the s-channel, because the coupling $e^{+}P^{+}H^{++}$ is proportional to electron mass, \cite{cnt1} in this work the couplings are interchanged, therefore we have not take into account such contributions.  Using the interaction Lagrangians
(\ref{pot}) and (\ref{yukt}) we evaluate the differential cross section for this reaction \par
\begin{eqnarray} 
\frac{d \hat{\sigma}}{d\cos \theta} & = & \frac{\beta \ \alpha^{2} \ \pi (\Lambda_{\gamma_\mu})^{2}}{8 s^{3}} \left [8 m_{\pm \pm}^{4} -8 m_{\pm \pm}^{2} (t+u)+ 16 m_{\pm \pm}^{2} \ m_{e}^{2} + 2 t^{2} \right. \nonumber \\
&&\left. + 4tu - 8 m_e^{2} (t+u) + 2 u^{2} - 2s^{2}+ 8 m_{e}^{2} \right] \nonumber \\
&&+\frac{\beta \ \alpha^{2} \ \pi (\Lambda_{Z(Z')\mu})^{2}}{32 s \sin^{2} \theta_{W} \cos^{2}\theta_{W} [(s- m_{Z,Z'}^{2})^{2} + m_{Z,Z'}^{2} \ \Gamma_{Z,Z'}^{2}]} \left[-8 m_{\pm\pm}^2 s[(g_{V}^{\ell(\ell^{'})})^{2} + (g_{A}^{\ell(\ell^{'})})^{2}] \right. \nonumber \\
&& \left. + 32 m_­{\pm \pm}^2 m_{e}^{2} (g_{A}^{\ell(\ell^{'})})^{2} - 2 t^{2} [(g_{V}^{\ell(\ell^{'})})^{2} + (g_{A}^{\ell(\ell^{'})})^{2}] + 4tu [(g_{V}^{\ell(\ell^{'})})^{2} + (g_{A}^{\ell(\ell^{'})})^{2}] \right. \nonumber \\
&& - 2u^{2} [(g_{V}^{\ell(\ell^{'})})^{2} + (g_{A}^{\ell(\ell^{'})})^{2}] \nonumber \\
&&\left. + 2 s^{2} [(g_{V}^{\ell(\ell^{'})})^{2} + (g_{A}^{\ell(\ell^{'})})^{2}] -8 s m_{e}^{2} (g_{A}^{\ell(\ell^{'})})^{2} \right ] \nonumber \\
&& +\frac{\beta \ m_{e}^{2} \ (\Lambda_1)^{2}}{128 \ \pi v_{W}^{2} \ s \ [(s- m_{H_{1}^{0}}^{2})^{2} + m_{H_{1}^{0}}^{2} \ \Gamma_{H_{1}^{0}}^{2}]} \left(s- 2 \ m_{e}^{2} \right) \nonumber \\
&& +\frac{\beta \ m_{e}^{2} \ v_{\rho}^{2} \ (\Lambda_2)^{2}}{128 \ \pi \ v_{W}^{2} \ \ v_{\eta}^{2} \ s \ [(s- m_{H_{2}^{0}}^{2})^{2} + m_{H_{2}^{0}}^{2} \ \Gamma_{H_{2}^{0}}^{2}]} \left(s- 2 \ m_{e}^{2} \right) \\
&& - \frac{\beta \ m_{e}^{2} \ v_{\rho} \ (\Lambda_1) \ (\Lambda_2) }{64 \ \pi \ v_{W}^{2} \ \ v_{\eta} \ s \ (s- m_{H_{1}^{0}}^{2} + im_{H_{1}^{0}} \ \Gamma_{H_{1}^{0}}) \ (s- m_{H_{2}^{0}}^{2} + im_{H_{2}^{0}} \ \Gamma_{H_{2}^{0}})} \left(s- 2 \ m_{e}^{2} \right) \nonumber \\
&&+\frac{\beta \ \alpha^{2} \ \pi \ (\Lambda_{\gamma_\mu}) \ (\Lambda_{Z\prime_\mu})}{8 \ \sin \theta_{W} \ cos \theta_{W} \ s (s- m_{Z'}^{2}) } \left[-8 \ m^{\pm\pm 2} s {g_{V}^{\ell^{'}}} - 2 t^{2} {g_{V}^{\ell^{'}}} \right. \nonumber \\
&&\left. + 4 t u {g_{V}^{\ell^{'}}} -2 u^{2} {g_{V}^{\ell^{'}}} + 2s^{2} {g_{V}^{\ell^{'}}} \right] \ . \nonumber \\
\end{eqnarray} 
The primes $\left(^\prime\right)$ is for the case when we take a boson 
$Z'$, $\Gamma_{Z,Z'}$ are the total width of the boson Z and $Z'$
\cite{cieton1,cieton2}, $\beta$ is the velocity of the Higgs in the center of mass of the process, $\alpha$ is the fine structure constant, which we take equal to $\alpha =1/128$, $g^{\ell}_{V, A}$ are
the standard coupling constants, $g^{\ell^\prime}_{V, A}$ are the 3-3-1 lepton coupling constants, $m_{Z}$ is the mass of the $Z$ boson,
$\sqrt{s}$ is the center of mass energy of the $e^{-} e^{+}$ system, $t 
= m_{\pm \pm}^{2} - (1 - \beta \cos \theta)s/2$ and $u = m_{\pm \pm}^{2} - (1 + \beta \cos \theta)s/2$, where $\theta$ is the angle between the heavy DCHBs and the incident electron, in the c. m. frame. For 
$Z^\prime$ boson we take $m_{Z^\prime} = \left(0.6 - 3\right)$ TeV,
since $m_{Z^\prime}$ is proportional to the VEV $v_\chi$  \cite{PP92}. For the standard model parameters we assume PDG values,  {\it i. e.}, 
$m_Z = 91.19$ GeV, $\sin^2{\theta_W} = 0.2315$ and $m_W = 80.33$ GeV
\cite{Cea98} and $v_{\eta}$, $v_{\rho}$ are the VEV.  The $\Lambda_i$, where i stands for $H_{1}^{0}, H_{2}^{0}$, are the vertex  strengths of these  bosons to $H^{--} H^{++}$, the $\Lambda_{\gamma_\mu}$ is the vertex strengths of the foton to $H^{--} H^{++}$, and the $\Lambda_{Z(Z')\mu}$ are of the bosons Z and $Z^{'}$ to $H^{--} H^{++}$. The analytical expressions for these vertex strengths are
\begin{eqnarray}
\Lambda_{\gamma_\mu} & = & \frac{v_{\chi}^{2}- v_{\eta}^{2}}{v_{\chi}^{2}+ v_{\eta}^{2}} ,    \\ 
\Lambda_{Z_\mu} & = &  -i \frac{(1- 4 \sin^{2} \theta_{W}) v_{\eta}^{2}+ 4 \sin^{2}  \theta_{W} \ v_{\chi}^{2}}{4 \ \sin \theta_{W} \ \cos \theta_{W} \ (v_{\chi}^{2}+ v_{\eta}^{2})} , \\ 
\Lambda_{Z^\prime_\mu} & = & -\frac{2(1-7 \sin^{2} \theta_{W}) v_{\chi}^{2}- (1- 10 \sin^{2} \theta_{W}) v_{\eta}^{2}}{4 \ \sin \theta_{W} \ \cos \theta_{W}
\sqrt{3(1- 4 \sin^{2} \theta_{W})} (v_{\chi}^{2}+ v_{\eta}^{2})} , \\ 
\Lambda_1 & = & -i (\frac{2[(2\lambda_{6}+ \lambda_{9}) v_{\eta}^{4}+ 2(2 \lambda_{2}+ \lambda_{9})v_{\eta}^{2} v_{\chi}^{2}+ 2(\lambda_{4} \ v_{\chi}^{2}+ \lambda_{5} \ v_{\eta}^{2}) v_{\rho}^{2}- f v_{\eta} v_{\rho} v_{\chi}]}{v_{W}(v_{\eta}^{2}+ v_{\chi}^{2})}), \\ 
\Lambda_2 & = & i (\frac{v_{\eta} [2 (-\lambda_{5}+  \lambda_{6}+ \lambda_{9}) v_{\eta}^{2} v_{\rho} + 2(2\lambda_{2}- \lambda_{4}+ \lambda_{9}) v_{\chi}^{2} \ v_{\rho}+ f v_{\eta} v_{\chi}]}{v_{W}(v_{\eta}^{2}+ v_{\chi}^{2})})   \  . 
\end{eqnarray}
The Higgs parameters $\lambda_i$ $\left(i = 1 \dots 9\right)$ must run from $-3$ to $+3$ in order to allow perturbative calculations. For $H_2^0$ we  take $m_{H_2^0} = \left( 0.2 - 3.0\right)$ TeV. It must be notice that here there is no contribution from the interference between the  scalar particle $H_{1(2)}^{0}$ and a vectorial one $\left(\gamma, Z {\mbox{ or }} Z^\prime\right)$ such as between the foton and the boson Z. \par
Referring to the signal, that is, $H^{\pm \pm} \to U^{\pm \pm} Z$,  it is necessary to compute the decay of $U^{\pm \pm}$, which total width into $J_{2,3} \ \bar{q}_{u,c,t} \ (\bar{J}_{2,3} \   q_{u,c,t})$ and $q_{d} \ \bar{J}_{1} \ (\bar{q}_{d} \ J_{1})$ quarks, $e^{\pm} P^{\pm}$ leptons, $\gamma H^{\pm \pm}$ foton and DCHBs, $Z(Z^\prime) H^{\pm \pm}$ gauge boson and DCHBs,  $H_{1}^{\pm} \ H_{2}^{\pm}$, $H_{1}^{0} H^{\pm \pm}$, $H_{2}^{0} H^{\pm \pm}$, $H_{3}^{0} H^{\pm \pm}$ and $h^{0} H^{\pm \pm}$ Higgs bosons, are  respectively given by
\begin{eqnarray}
\Gamma \left(U^{\pm \pm} \to {\rm all}\right)&=& \Gamma_{U^{\pm \pm} \to J_{2,3} \ \bar{q}_{u,c,t} \ (\bar{J}_{2,3} \   q_{u,c,t})} +  \Gamma_{U^{\pm \pm} \to q_{d} \ \bar{J}_{1} \ (\bar{q}_{d} \ J_{1}) }  + \Gamma_{U^{\pm \pm} \to \ell^{\pm} P^{\pm}}+ \Gamma_{U^{\pm \pm} \to \gamma \ H^{\pm \pm} } \nonumber  \\
&& + \Gamma_{U^{\pm \pm} \to Z \ H^{\pm \pm}} + \Gamma_{U^{\pm \pm} \to Z^\prime \ H^{\pm \pm}} + \Gamma_{U^{\pm \pm} \to H_{1}^{\pm} \ H_{2}^{\pm}} + \Gamma_{U^{\pm \pm} \to H_{1}^{0} \ H^{\pm \pm}} \nonumber  \\
&& + \Gamma_{U^{\pm \pm} \to H_{2}^{0} \ H^{\pm \pm}}+ \Gamma_{U^{\pm \pm} \to H_{3}^{0} \ H^{\pm \pm}} + \Gamma_{U^{\pm \pm} \to h^{0} \ H^{\pm \pm}}   \nonumber  , 
\end{eqnarray}
where we have for each the widths given above that
\begin{mathletters}  
\begin{eqnarray}  
\Gamma_{U^{\mp \mp} \to J_{2,3} \ \bar{q}_{u,c,t} \ (\bar{J}_{2,3} \   q_{u,c,t})} & = & \frac{\sqrt{1- \left ( \frac{m_{J_{2,3}}+ m_{\bar{q}_{u,c,t}}}{m_{U^{\mp \mp}}} \right )^{2}} \sqrt{1- \left ( \frac{m_{J_{2,3}}- m_{\bar{q}_{u,c,t}}}{m_{U^{\mp \mp}}} \right )^{2}}} {48 \pi m_{U^{\mp \mp}}}   (\Lambda_{UJq_u}^{2})  \nonumber  \\
&&\left (8 m_{U^{\pm \pm}}^{2}- 4m_{J_{2,3}}^{2}- 4m_{\bar{q}_{u,c,t}}^{2} \right)  ,  \\
\Gamma_{U^{\mp \mp} \to q_{d} \ \bar{J}_{1} \ (\bar{q}_{d} \ J_{1})} & = & \frac{\sqrt{1- \left ( \frac{m_{q_{d}}+ m_{\bar{J_{1}}}}{m_{U^{\mp \mp}}} \right )^{2}} \sqrt{1- \left ( \frac{m_{q_{d}}- m_{\bar{J}_{1}}}{m_{U^{\mp \mp}}} \right )^{2}}}{48 \pi m_{U^{\mp \mp}}}   (\Lambda_{UJq_d}^{2})  \nonumber  \\
&&\left (8 m_{U^{\mp \mp}}^{2}- 4m_{\bar{J}_{1}}^{2}- 4m_{q_{d}}^{2} \right)  ,  \\
\Gamma_{U^{\mp \mp} \to \ell^{-} \ P^{-} \ (\ell^{+} \ P^{+})} & = & \frac{\sqrt{1- \left ( \frac{m_{\ell}+ m_{P}}{m_{U^{\mp \mp}}} \right )^{2}} \sqrt{1- \left ( \frac{m_{\ell}- m_{P}}{m_{U^{\mp \mp}}} \right )^{2}}}{48 \pi m_{U^{\mp \mp}}}   (\Lambda_{UlP}^{2})  \nonumber  \\
&&\left (8 m_{U^{\mp \mp}}^{2}- 4m_{\ell}^{2}- 4m_{P}^{2} \right)  ,  \\
\Gamma_{U^{\mp \mp} \to \gamma \ H^{\mp \mp}} & = & \frac{\left (1- \frac{m_{\mp \mp}^{2}}{m_{U^{\mp \mp}}^{2}} \right)}{16 \pi m_{U^{\mp \mp}}} [(\Lambda_{UAH^{++}})^{\mu \nu}(\Lambda_{UAH^{++}})_{\mu \nu}]^{2}  ,  \\
\Gamma_{U^{\mp \mp} \to Z \ H^{\mp \mp}} & = & \frac{\sqrt{1- \left ( \frac{m_{Z}+ m_{\mp \mp}}{m_{U^{\mp \mp}}} \right )^{2}} \sqrt{1- \left ( \frac{m_{Z}- m_{\mp \mp}}{m_{U^{\mp \mp}}} \right )^{2}}}{48 \pi m_{U^{\mp \mp}}}  \left(2+ \frac{m_{U^{\mp \mp}}^{2}}{4 m_{Z}^{2}} \right)   \nonumber  \\
&&[(\Lambda_{UZH^{+ +}})^{\mu \nu}(\Lambda_{UZH^{+ +}})_{\mu \nu}]^{2}  ,  \\
\Gamma_{U^{\mp \mp} \to Z' \ H^{\mp \mp}} & = & \frac{\sqrt{1- \left ( \frac{m_{Z'}+ m_{\mp \mp}}{m_{U^{\mp \mp}}} \right )^{2}} \sqrt{1- \left ( \frac{m_{Z'}- m_{\mp \mp}}{m_{U^{\mp \mp}}} \right )^{2}}}{48 \pi m_{U^{\mp \mp}}}  \left(2+ \frac{m_{U^{\mp \mp}}^{2}}{4 m_{Z'}^{2}} \right)   \nonumber   \\
&& [(\Lambda_{UZLH^{+ +}})^{\mu \nu}(\Lambda_{UZLH^{+ +}})_{\mu \nu}]^{2}  ,  \\
\Gamma_{U^{\mp \mp} \to H_{1}^{-} \ H_{2}^{-}} & = & \frac{\sqrt{1- \left ( \frac{m_{H_{1}^{-}}+ m_{H_{2}^{-}}}{m_{U^{\mp \mp}}} \right )^{2}} \sqrt{1- \left ( \frac{m_{H_{1}^{-}}- m_{H_{2}^{-}}}{m_{U^{\mp \mp}}} \right )^{2}}}{48 \pi m_{U^{\mp \mp}}}   \nonumber   \\
&&\left(m_{U}^{2}- 2m_{H_{1}^{-}}^{2}- 2m_{H_{2}^{-}}^{2}  \right) (\Lambda_{UH_{1}H_{2}}^{2})  ,  \\
\Gamma_{U^{\mp \mp} \to H_{i}^{0} \ H^{\mp \mp}} & = & \frac{\sqrt{1- \left ( \frac{m_{H_{i}^{0}}+ m_{\mp \mp}}{m_{U^{\mp \mp}}} \right )^{2}} \sqrt{1- \left ( \frac{m_{H_{i}^{0}}- m_{\mp \mp}}{m_{U^{\mp \mp}}} \right )^{2}}}{48 \pi m_{U^{\mp \mp}}}   \nonumber   \\
&&\left(m_{U}^{2}- 2m_{H_{i}^{0}}^{2}- 2m_{\mp \mp}^{2}  \right) (\Lambda_{UH^0_{i}H^{+ +}}^{2})  ,
\label{xx}   
\end{eqnarray}
\end{mathletters}\noindent
where $H_{i}^{0}$ denote the $H_{1}^{0}$, $H_{2}^{0}$, $H_{3}^{0}$ and  $h^{0}$ and the couplings are given by
\begin{eqnarray}
\Lambda_{UJqu(UJqd)} & = & -i\frac{e}{2 \sqrt{2} \sin \theta_{W}} ,    \\ 
\Lambda_{UlP} & = &  -i\frac{e}{2 \sqrt{2} \sin \theta_{W}} ,    \\ 
(\Lambda_{UAH^{+ +}})_{\mu \nu} & = & i g_{\mu \nu} \ \frac{\sqrt{2} \ v_{\eta} v_{\chi}}{\sin_{\theta_{W}} \sqrt{v_{\eta}^{2}+ v_{\chi}^{2}}}   , \\ 
(\Lambda_{UZH^{+ +}})_{\mu \nu} & = & i g_{\mu \nu} \ \frac{\sqrt{2} \ v_{\eta} v_{\chi}}{\cos_{\theta_{W}} \sqrt{v_{\eta}^{2}+ v_{\chi}^{2}}}  , \\
(\Lambda_{UZLH^{+ +}})_{\mu \nu} & = & - i g_{\mu \nu} \frac{\sqrt{2 (1-4 \sin_{{\theta}_{W}}^{2})} \ v_{\eta} v_{\chi}}{\sin_{\theta_{W}} \cos_{\theta_{W}}  \sqrt{3 (v_{\eta}^{2}+ v_{\chi}^{2})}} , \\
\Lambda_{UH_{1}H_{2}} & = & - \frac{v_{\eta} v_{\chi}}{2\sqrt{2} \sin_{\theta_{W}} v_{W}(v_{\rho}^{2}+ v_{\chi}^{2})} , \\
\Lambda_{UH_{1}^0H^{+ +}} & = & - \frac{v_{\eta} v_{\chi}}{2\sqrt{2} \sin_{\theta_{W}} v_{W}  \sqrt{v_{\eta}^{2}+ v_{\chi}^{2}}}  , \\
\Lambda_{UH_{2}^0H^{+ +}} & = &  \frac{v_{\rho} v_{\chi}}{2\sqrt{2} \sin_{\theta_{W}} v_{W}  \sqrt{v_{\eta}^{2}+ v_{\chi}^{2}}}  , \\
\Lambda_{UH_{3}^0H^{+ +}} & = & - \frac{v_{\eta}}{2 \sqrt{2} \sin_{\theta_{W}}   \sqrt{v_{\eta}^{2}+ v_{\chi}^{2}}}  , \\
\Lambda_{Uh^{0}H^{+ +}} & = & i \frac{v_{\eta}}{2\sqrt{2} \sin_{\theta_{W}}   \sqrt{v_{\eta}^{2}+ v_{\chi}^{2}}}  \  .
\end{eqnarray}
\section{RESULTS AND CONCLUSIONS}
In the following we present the cross section for the process $e^{+} e^{-} \rightarrow H^{\pm \pm} H^{\mp \mp}$ for
the ILC (1.5 TeV) and CLIC (3 TeV), where we have chosen for the parameters, masses and the VEV, the following representative values: $\lambda_{1} =-1.2$,  $\lambda_{2}=\lambda_{3}=-\lambda_{6}=\lambda_{8}=-1$, $\lambda_{4}= 2.98$ $\lambda_{5}=-1.57$, $\lambda_{7}=-2$,  $\lambda_{9}=-0.8$, $v_{\eta}=195$ GeV  and with other particles masses as given in Table I, it is to notice that the value of $\lambda_{9}$ was chosen this way in order to guarantee the approximation $-f \simeq v_{\chi}$, \cite{cnt2,ton1} and because the masses of $m_{h^0}$, $m_{H_1^{\pm}}$ and $m_{H_{2}^{\pm}}$  depend on the parameter $f$ and therefore they can not be fixed by any value of $v_{\chi}$,  \cite{cnt1,cnt2}, so when we have $m_{\pm \pm}= 500(700)$ GeV, $v_{\chi}=1000$ GeV, the masses of $H_{2}^{\pm}$ and $h$ are  $m_{H_{2}}^{\pm}= 671.9(917.1)$ GeV, and $m_h = 1756.2(1017.6)$ GeV, and in the case of $v_{\chi}=1500$ for $m_{\pm \pm}= 500(700)$ GeV, the values of the mass of $H_{2}^{\pm}$ and $h$ are $m_{H_{2}}^{\pm}= 901.6(1223.6)$ GeV and $m_h = 2802.7(2052.2)$ GeV, respectively.\par
The mass of $m_{Z^{\prime}}$ taken in Table I is in accord with the estimates of the CDF and D0 experiments, which probes the $Z^{\prime}$ masses in the 500-800 GeV range, \cite{tait}, while the reach of the LHC is superior for higher masses, that is 1 TeV $< m_{Z^{\prime}} \leq 5$ TeV, \cite{freitas}, concerning to Higgs the LHC is able to discover the Higgs boson with a mass up to TeV's and to check its basic properties.
\begin{widetext}
\begin{center}
\begin{table}
\caption{\label{tab1} \footnotesize\baselineskip = 12pt 
Values of the masses for $v_\eta = 195$ GeV and the sets of parameters given in the text. All the values in this table are given in GeV.}
\begin{tabular}{ccccccccccccc}  
\hline\hline
$v_\chi$ & $m_E$ & $m_M$ & $m_T$ & $m_{H^0_2}$ & $m_{H^0_3}$ & $m_V$ & $m_U$ & $m_{Z^\prime}$ & $m_{J_1}$ & $m_{J_2}$ & $m_{J_3}$\\ 
1000 & 148.9 & 875 & 2000 & 1017.2 & 2000 & 467.5 & 464 & 1707.6 & 1000 & 1410 & 1410 \\
1500 & 223.3 & 1312.5 & 3000 & 1525.8 & 3000 & 694.1 & 691.8 & 2561.3 & 1500 & 2115 & 2115 \\
\hline\hline
\end{tabular}
\end{table}
\end{center}
\end{widetext}

\begin{figure}  
\includegraphics [scale=.67]{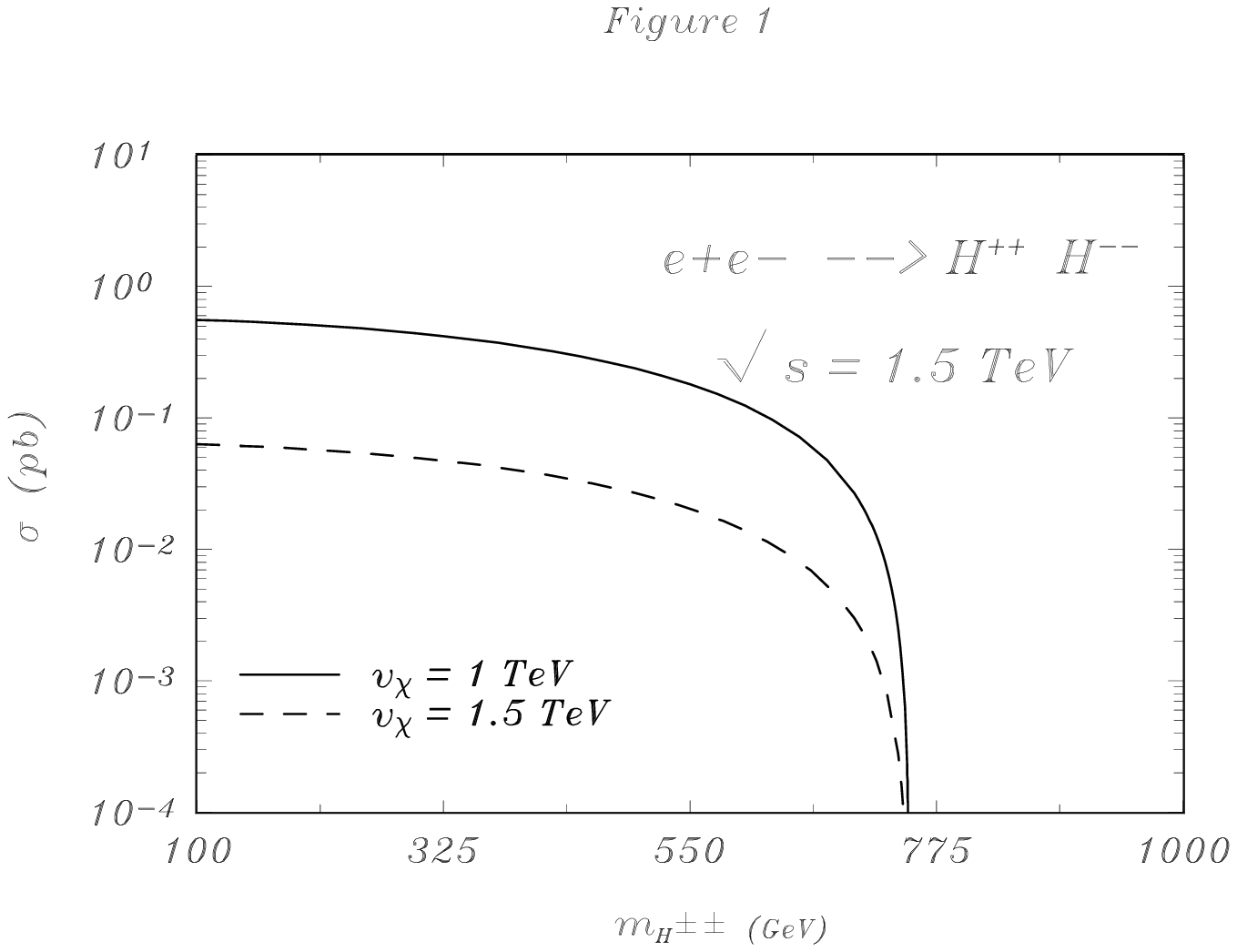}
\caption{ \label{fig1}  Total cross section for the process $e^{-} e^{+} \rightarrow H^{--} H^{++}$ as function of $m_{\pm \pm}$ at $\sqrt{s}$ = $1500$ GeV:  (a) $v_{\chi}$ = $1$ TeV (solid line), and  (b) $v_{\chi}$ = $1.5$ TeV (dashed line).}
\end{figure}  

\begin{figure}  
\includegraphics [scale=.67]{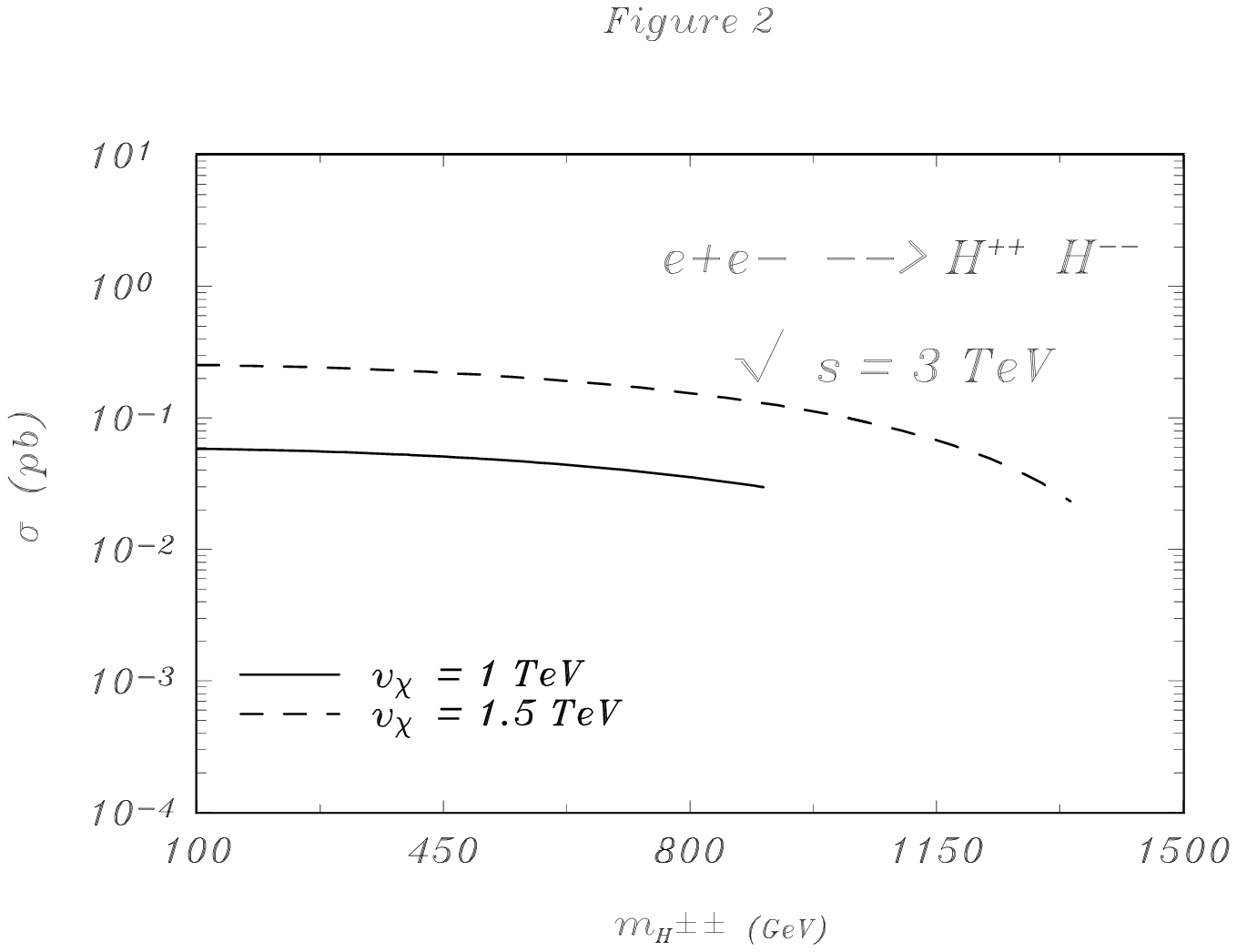}
\caption{ \label{fig2} Total cross section for the process $e^{-} e^{+} \rightarrow H^{--} H^{++}$ as function of $m_{\pm \pm}$ at $\sqrt{s} = 3$ TeV: (a) $v_{\chi} = 1$ TeV (solid line), and  (b) $1.5$ TeV (dashed line).}
\end{figure}  

\begin{figure}  
\includegraphics [scale=.67]{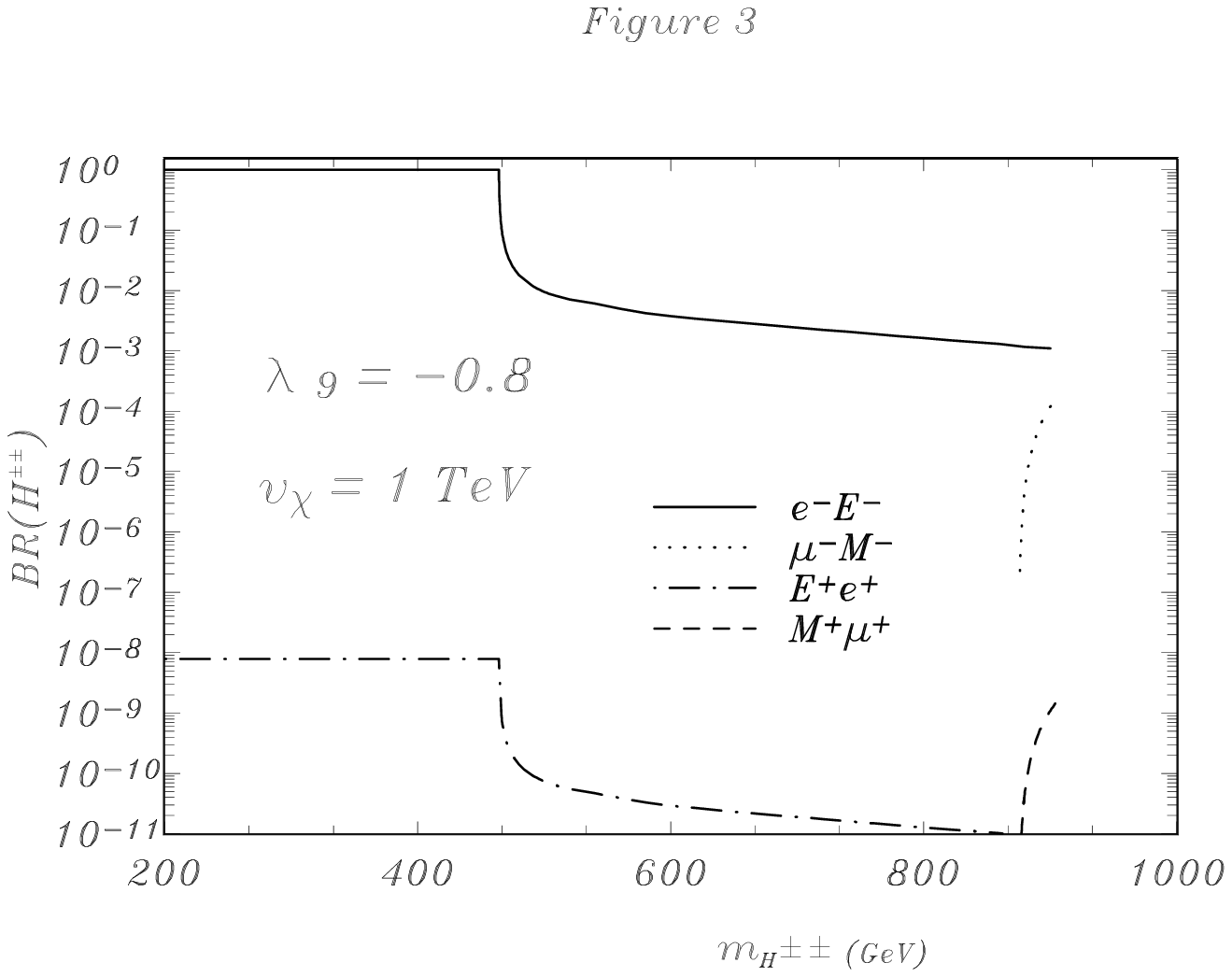}
\caption{ \label{fig3} Branching ratios for the DCHBs decays as functions of $m_{\pm \pm}$ for $\lambda_{9}=- 0.8$, $v_{\chi}=1$ TeV for the leptonic sector.}
\end{figure}  

\begin{figure}  
\includegraphics [scale=.67]{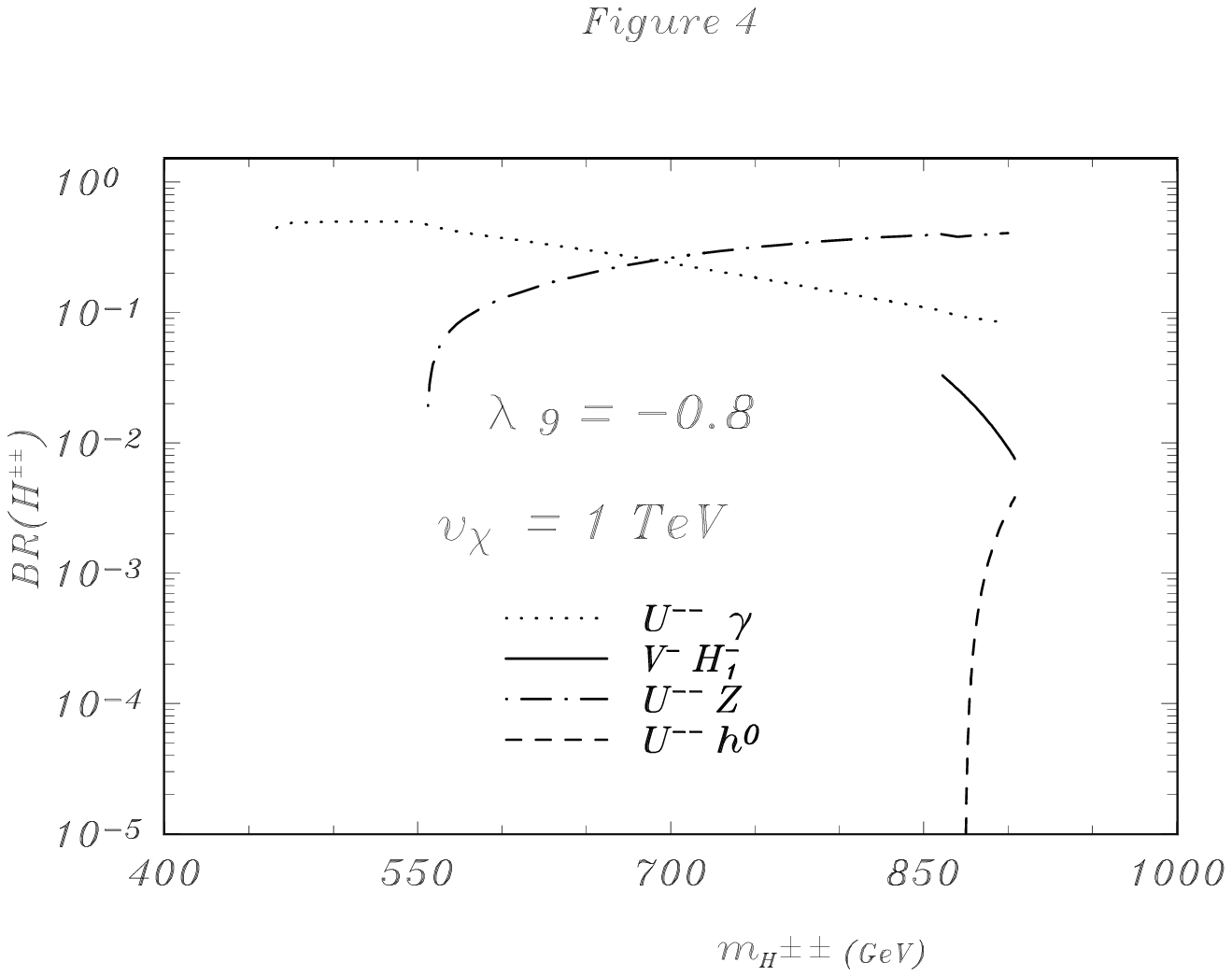}
\caption{ \label{fig4} Branching ratios for the DCHBs decays as functions of $m_{\pm \pm}$ for $\lambda_{9}=- 0.8$, $v_{\chi}=1$ TeV for the bosonic sector.}
\end{figure}  

\begin{figure}  
\includegraphics [scale=.67]{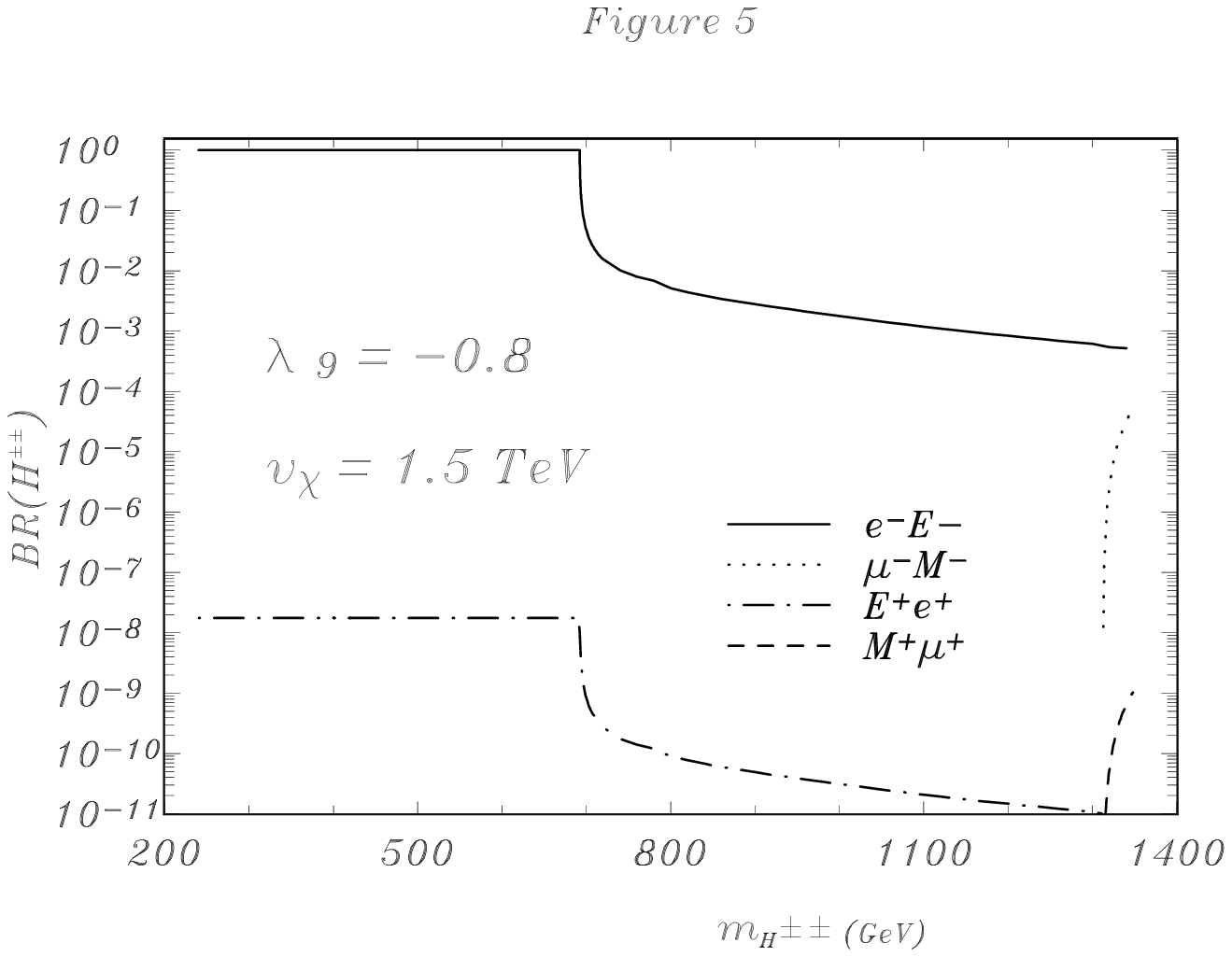}
\caption{ \label{fig5} Branching ratios for the DCHBs decays as functions of $m_{\pm \pm}$ for $\lambda_{9}=- 0.8$, $v_{\chi}=1.5$ TeV for the leptonic sector.}
\end{figure}  

In Figs. 1 and 2, we show the cross section $e^{+} e^{-} \rightarrow  H^{\pm \pm} H^{\mp \mp}$, these processes will be studied in two cases, the one where we put for the VEV $v_{\chi}=1000$ GeV and the other $v_\chi = 1500$ GeV, respectively. Considering that the expected integrated luminosity for both colliders will be of order of $3.8 \times 10^5$ pb$^{-1}$/yr and $3 \times 10^6$ pb$^{-1}$/yr respectivelly, then the statistics we are expecting are the following.\par
The ILC collider gives a total of $ \simeq 9.1 \times 10^4$ events per year, if we take the mass of the boson $m_{\pm \pm}= 500$ GeV, $v_{\chi}=1000$ GeV. Considering that the signal for $H^{\mp \mp}$ are $U^{--} \gamma$ and $U^{++} \gamma$ and taking into account that the branching ratios for these  particles would be $BR(H^{--} \to U^{--} \gamma) = 49.5 \%$ and $BR(H^{++} \to U^{++} \gamma) = 49.5 \%$, see Figs. 3 and 4, for the mass of the Higgs boson $m_{\pm \pm}= 500$ GeV, $v_{\chi}=1000$ GeV, and that the particles $U^{\mp \mp}$ decay into $e^{-} P^{-}$ and $e^{+} P^{+}$, whose branching ratios for these particles would be $BR(U^{--} \to e^{-} P^{-}) = 50 \% $ and $BR(U^{++} \to e^{+} P^{+}) = 50 \%$, see Figs. 7 and 8, then  we would have approximately $\simeq 5.6 \times 10^{3}$ events per year for the ILC, regarding the VEV $v_{\chi}=1500$ GeV it  will not give any event because it its restricted by the values of $m_{U^{\pm \pm}}$ which in this case give $m_{U^{\pm \pm}}=691.8$ GeV, see Table I. 
\begin{center}
\begin{table}
\caption{\label{tab2}\footnotesize\baselineskip = 12pt Branching ratios from $H^{\pm\pm} \to e^{\pm\pm}$ and events per year (total and partial), for $v_\chi$ = $1000$ and $1500$ in function of $m_{++}$, $m^{\pm}_{H_2}$ and $m_h$, which their units and the unit of $v_\chi$ are in GeV.}
\begin{tabular}{|c|c|c|c|c|c|c||c|c|c|}
\hline\hline
& & & & BR in $\%$ for &NLC &NLC & CLIC & CLIC \\
\raisebox{1.8ex}{$v_\chi$} & \raisebox{1.8ex}{$m_{++}$} & \raisebox{1.8ex}{$m^{\pm}_{H_2}$} & \raisebox{1.8ex}{$m_h$} & $H^{\pm \pm} \to U^{\pm \pm}\gamma$ & Total ev/yr & BR ev/yr & Total ev/yr & BR ev/yr \\
\hline \hline
\raisebox{-1.5ex}{$1000$} &
$500$ & 671.9 & 1756.2 & 49.5 & 9.1 $\times$ $10^4$ & 5.6 $\times$ $10^3$ & 1.5 $\times$ $10^5$ & 9.1 $\times$ $10^3$\\
& $700$ & 917.1 & 1017.6 & 23.7 & 1.0 $\times$ $10^4$ & 140 & 1.2 $\times$ $10^5$ & 1.7 $\times$ $10^3$\\
\hline
\raisebox{-1.5ex}{$1500$} &
$500$ & 901.6 & 2802.7 & - & - & - & - & - \\
& $700$ & 1223.6 & 2052.2 & 47.7 & 1.1 $\times$ $10^{3}$ & 63 & 5.4 $\times$ $10^5$ & 3.0 $\times$ $10^4$ \\
\hline\hline
\end{tabular}
\end{table}
\end{center}
\begin{figure}  
\includegraphics [scale=.67]{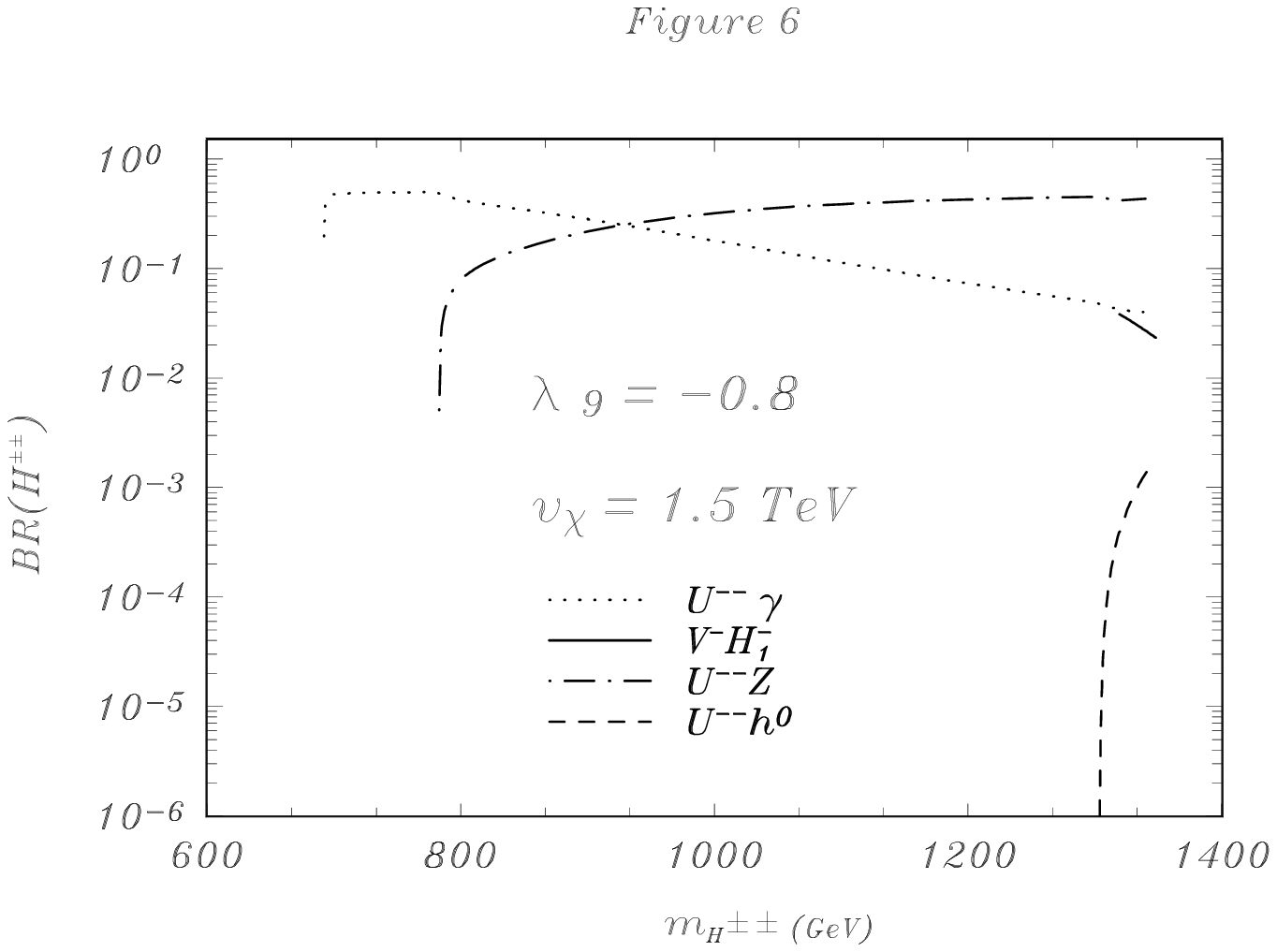}
\caption{ \label{fig6} Branching ratios for the DCHBs decays as functions of $m_{\pm \pm}$ for $\lambda_{9}=- 0.8$, $v_{\chi}=1.5$ TeV for the bosonic sector.}
\end{figure}  
\begin{figure}  
\includegraphics [scale=.67]{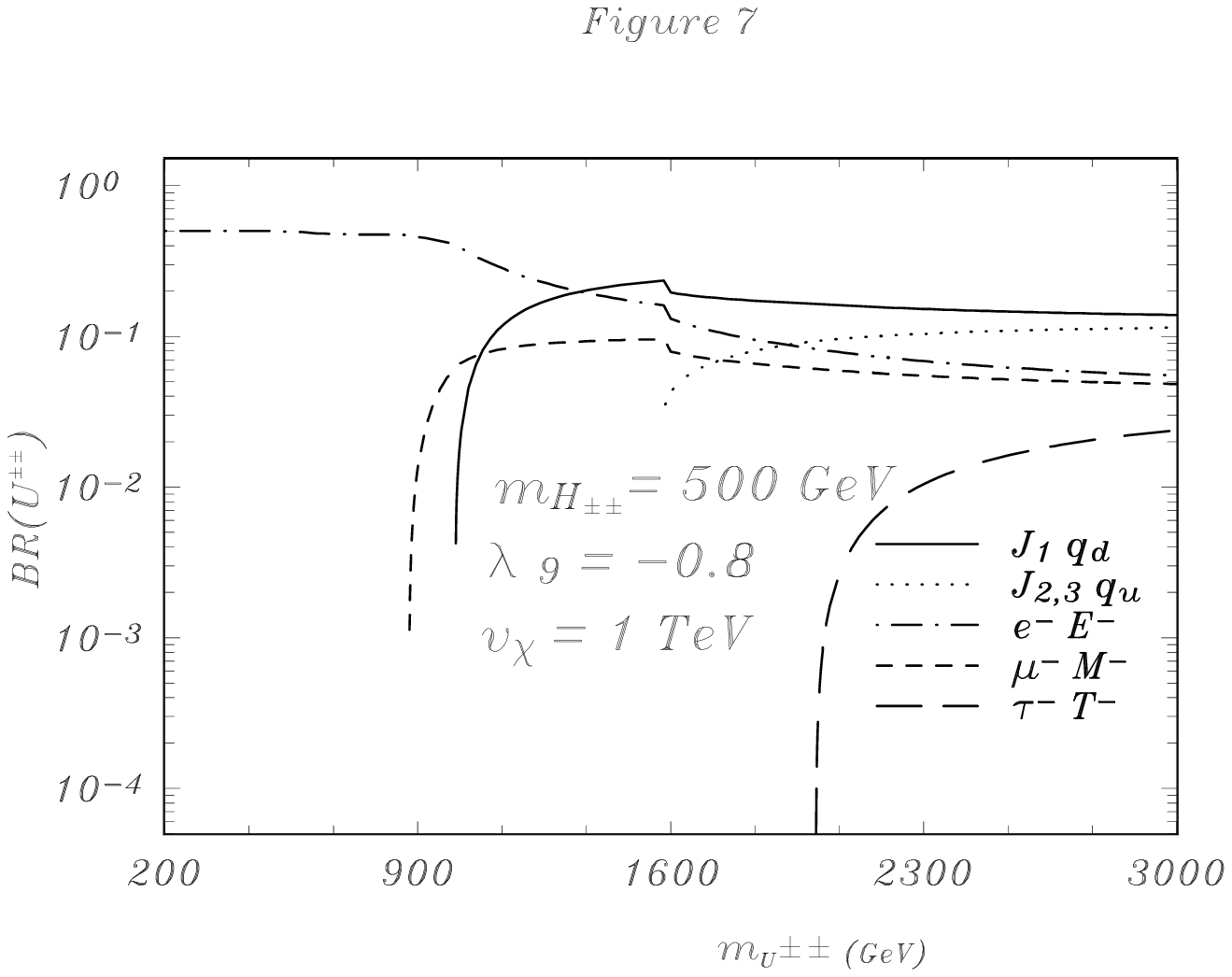}
\caption{ \label{fig7} Branching ratios for the doubly charged gauge bosons decays as functions of $m_{U^{\pm \pm}}$ for $\lambda_{9}=- 0.8$, $v_{\chi}=1$ TeV and $m_{\pm \pm}=500$ GeV for the leptonic sector.}
\label{fig:7}
\end{figure}  
\begin{figure}  
\includegraphics [scale=.67]{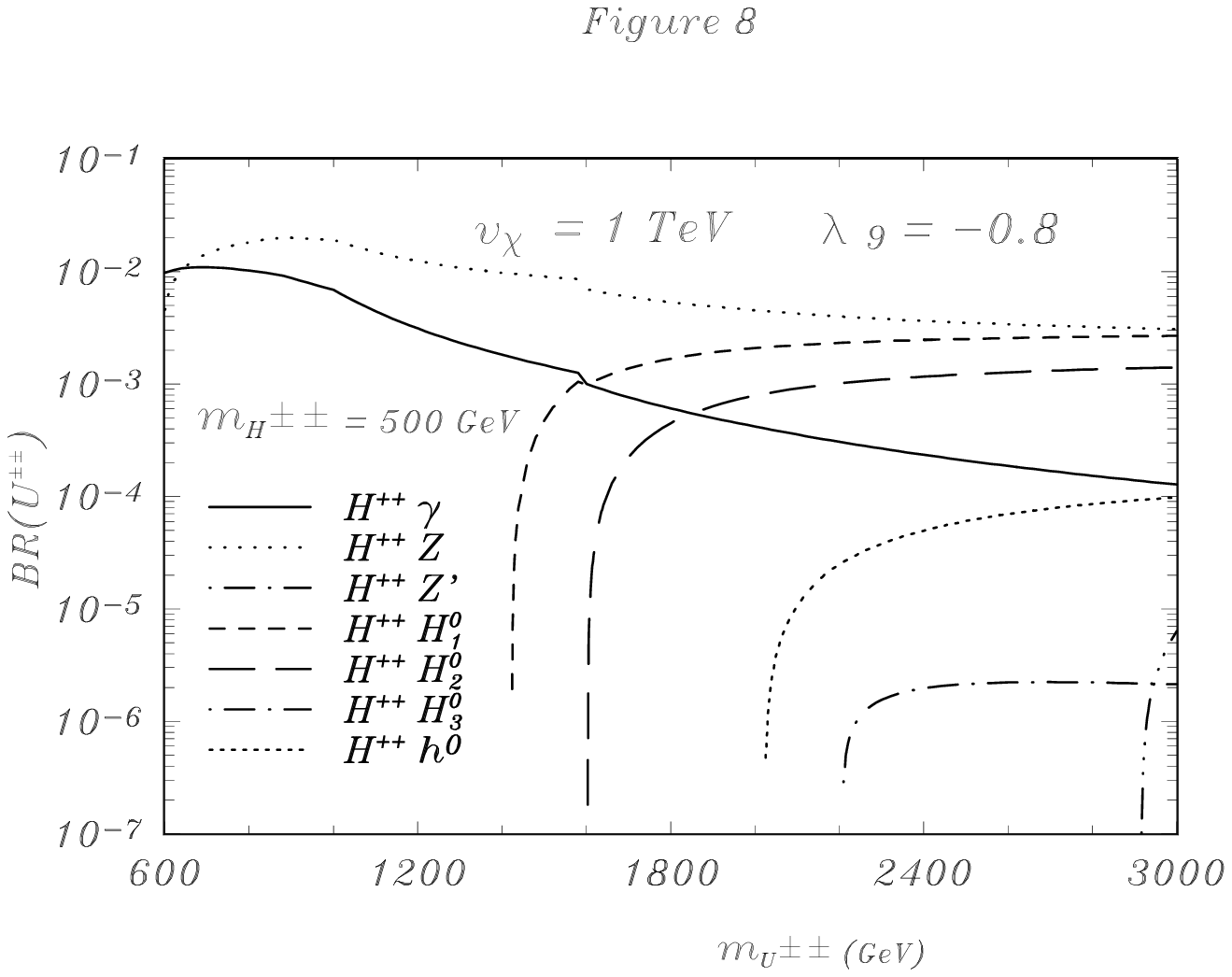}
\caption{ \label{fig8} Branching ratios for the doubly charged gauge bosons decays as functions of $m_{U^{\pm \pm}}$ for $\lambda_{9}=- 0.8$, $v_{\chi}=1$ TeV and $m_{\pm \pm}=500$ for the bosonic sector.}
\end{figure} 
\begin{figure}  
\includegraphics [scale=.67]{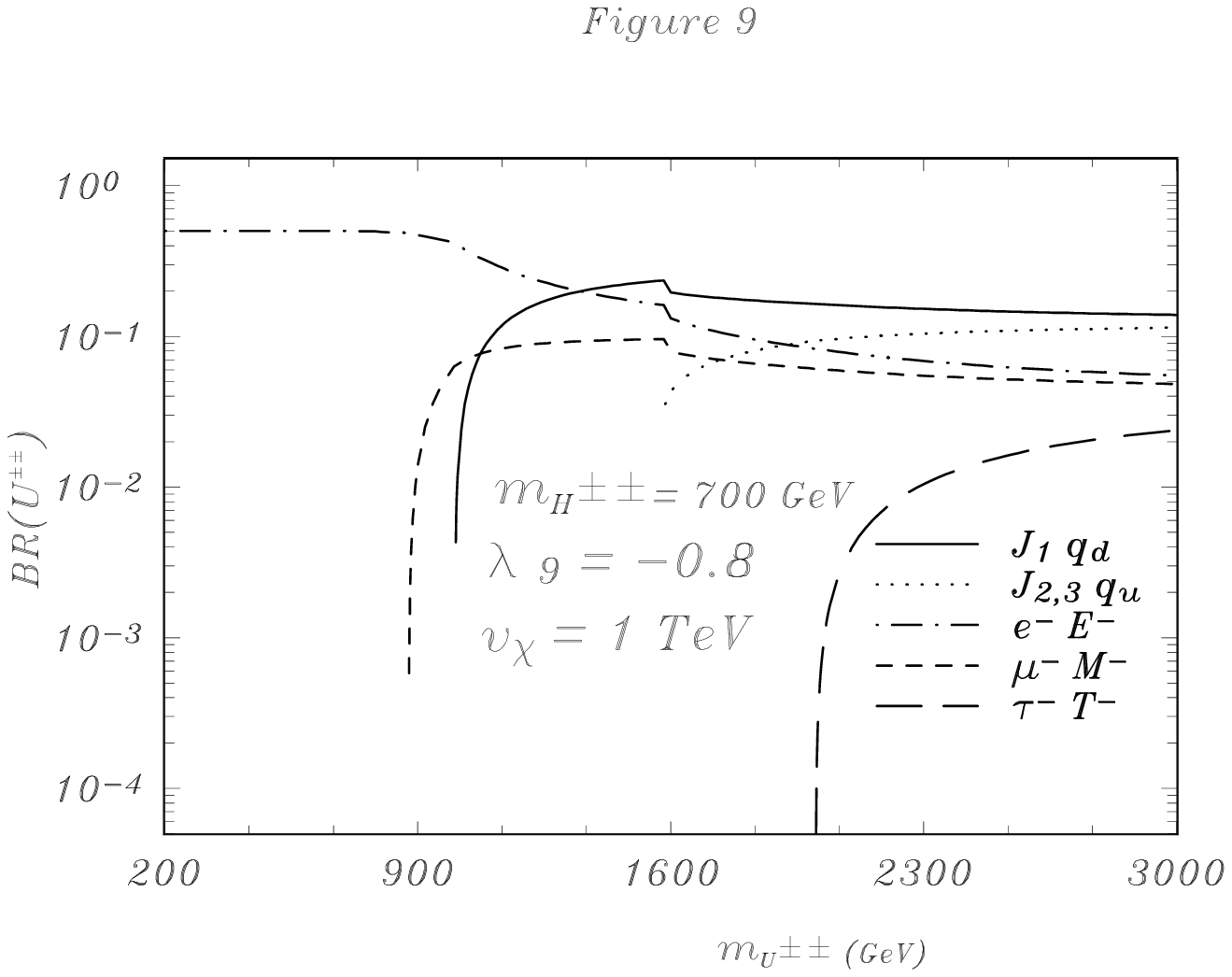}
\caption{ \label{fig9} Branching ratios for the doubly charged gauge bosons decays as functions of $m_{U^{\pm \pm}}$ for $\lambda_{9}=- 0.8$, $v_{\chi}=1$ TeV and $m_{\pm \pm}=700$ GeV for the leptonic sector.}
\end{figure}  
\begin{figure}  
\includegraphics [scale=.67]{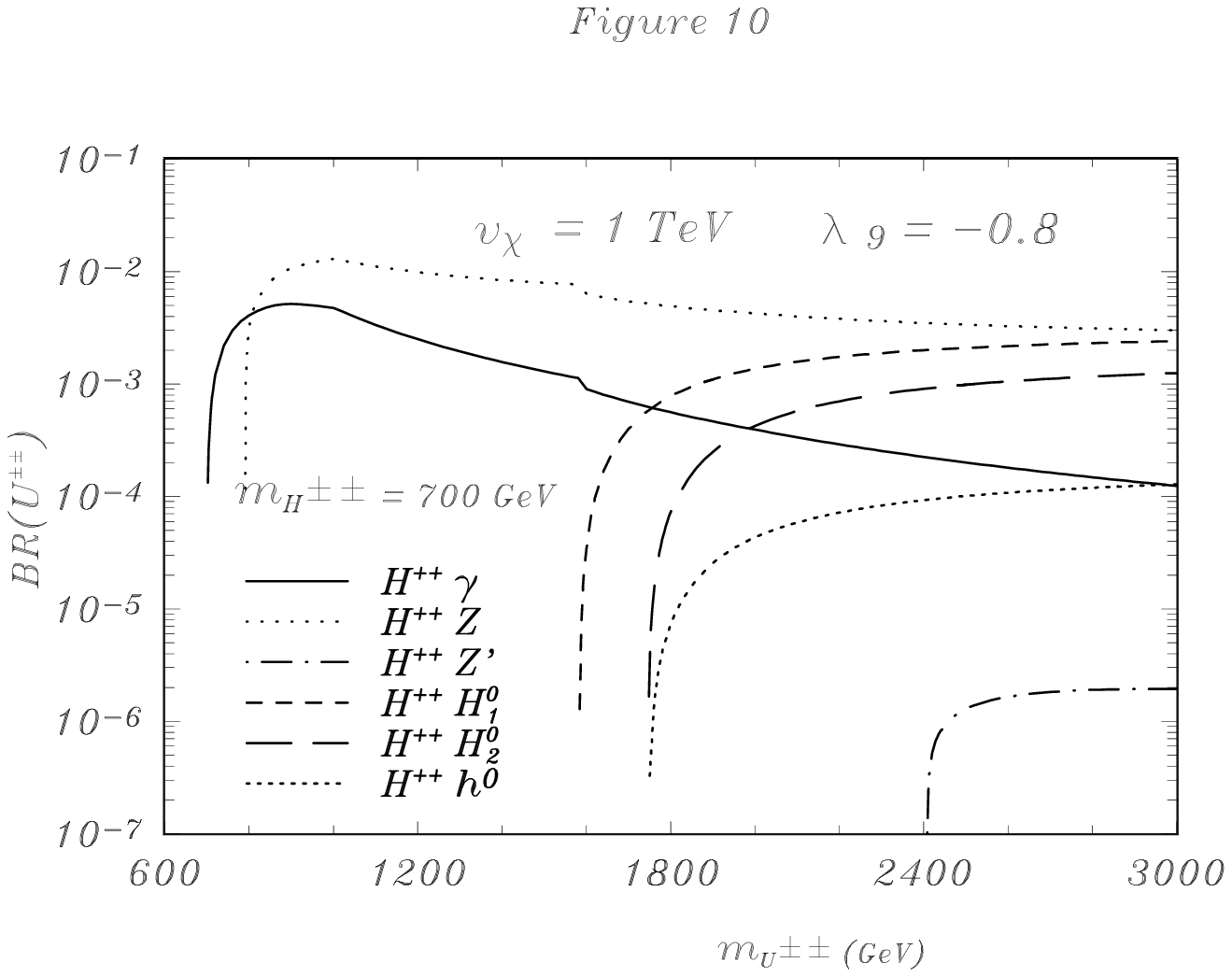}
\caption{ \label{fig10} Branching ratios for the doubly charged gauge bosons decays as functions of $m_{U^{\pm \pm}}$ for $\lambda_{9}=- 0.8$, $v_{\chi}=1$ TeV and $m_{\pm \pm}=700$ for the bosonic sector.}
\end{figure}
\begin{figure}  
\includegraphics [scale=.67]{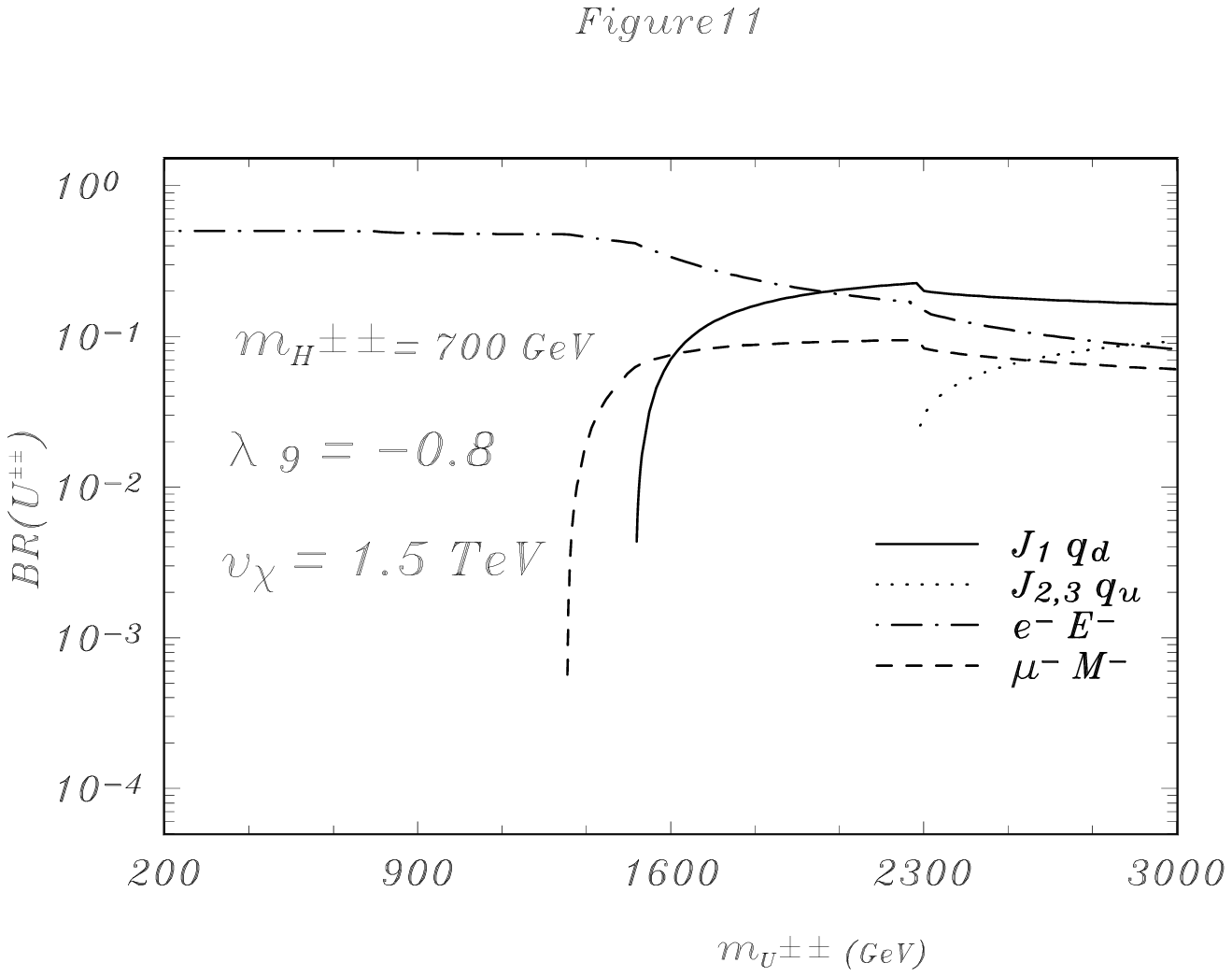}
\caption{ \label{fig11} Branching ratios for the doubly charged gauge bosons decays as functions of $m_{U^{\pm \pm}}$ for $\lambda_{9}=- 0.8$, $v_{\chi}=1.5$ TeV and $m_{\pm \pm}=700$ GeV for the leptonic sector.}
\end{figure}  
\begin{figure}  
\includegraphics [scale=.67]{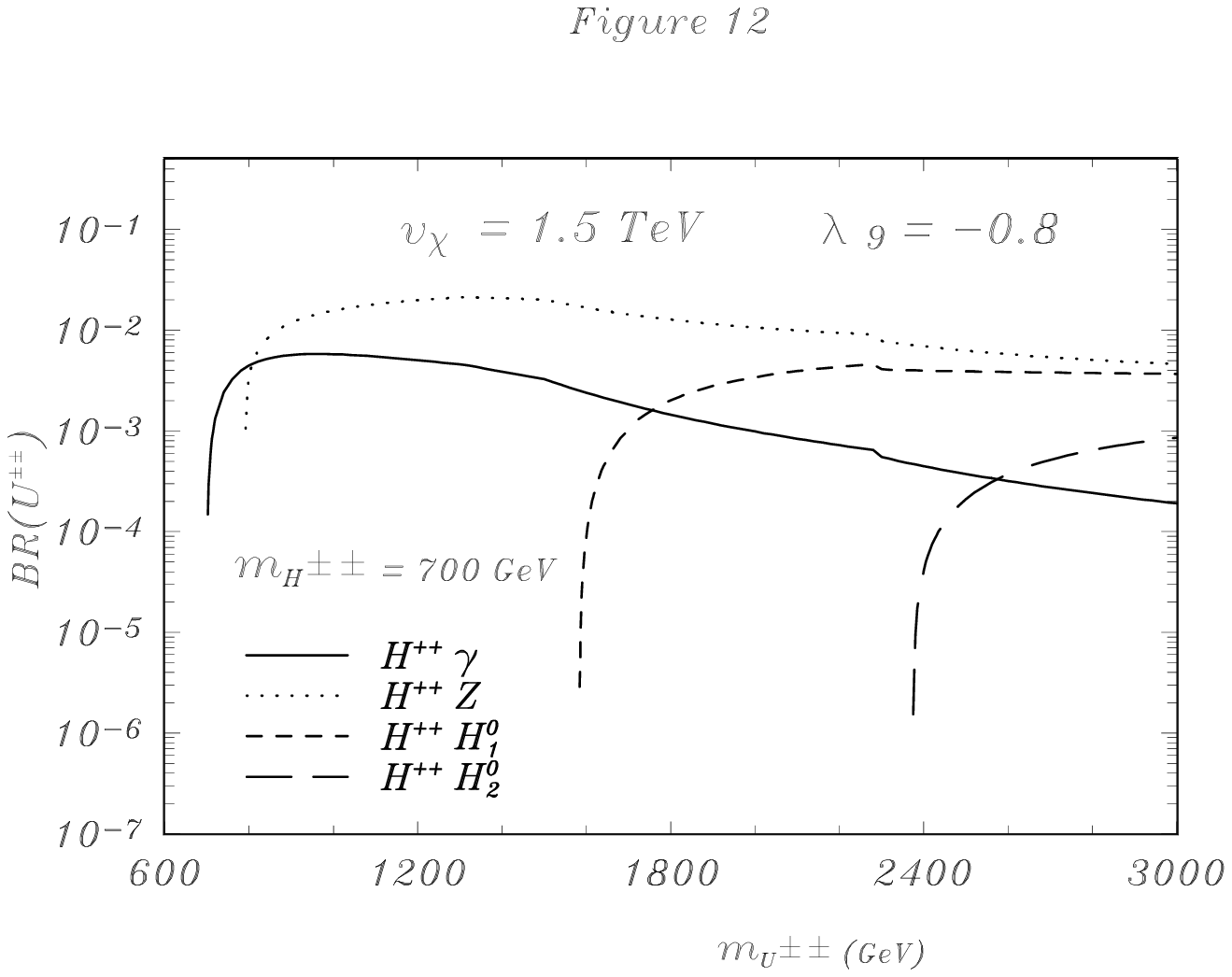}
\caption{ \label{fig12} Branching ratios for the doubly charged gauge bosons decays as functions of $m_{U^{\pm \pm}}$ for $\lambda_{9}=- 0.8$, $v_{\chi}=1.5$ TeV and $m_{\pm \pm}=700$ for the bosonic sector.}
\end{figure}
Taking now the mass of the Higgs boson $m_{\pm \pm}= 700$ GeV and the VEV value $v_{\chi}=1000$, we then have a total of $\simeq 10^4$ events per year. Considering now the same signal as above, whose branching  ratios  are equal to $BR(H^{--} \to U^{--} \gamma) = 23.7 \%$ and $BR(H^{++} \to U^{++} \gamma) = 23.7 \%$, see Figs. 3 and 4, for the mass of the Higgs boson $m_{\pm \pm}= 700$ GeV, $v_{\chi}=1000$ GeV, and that the particles $U^{\mp \mp}$ decay into $e^{-} P^{-}$ and $e^{+} P^{+}$, whose branching ratios for these particles would be $BR(U^{--} \to e^{-} P^{-}) = 50 \% $ and $BR(U^{++} \to e^{+} P^{+}) = 50 \%$, see Figs. 9 and 10, then  we would have approximately $\simeq 140$ events per year for the ILC, regarding the VEV $v_{\chi}=1500$ GeV  then we will have around $\simeq 1.1 \times 10^{3}$ events per year and   considering the same parameters signals as above, that is, $BR(H^{--} \to U^{--} \gamma) = 47.7 \%$ and $BR(H^{++} \to U^{++} \gamma) = 47.7 \%$, see Figs. 5 and 6, and $BR(U^{--} \to e^{-} P^{-}) = 50 \% $ and $BR(U^{++} \to e^{+} P^{+}) = 50 \%$, see Figs. 11 and 12, then  we would have approximately $\simeq 63$ events per year for the ILC. These results are showed in the Table II.\par 
The cross section for the CLIC collider is restricted by the mass of the DCHBs, because for $v_{\chi}=1000(1500)$ and $\lambda_{9}=-0.8$ the acceptable masses are up to $m_{\pm \pm} \simeq 903 (1346)$ GeV, see \cite{cnt2}, taking this into account we obtain a total of $ \simeq 1.5 \times 10^5$ events per year, if we take the mass of the boson $m_{\pm \pm}= 500$ GeV, $v_{\chi}=1000$ GeV. Considering the same signal as above for $H^{\mp \mp}$ production, that are, $U^{--} \gamma$ and $U^{++} \gamma$ and taking into account that the branching ratios for these  particles would be $BR(H^{--} \to U^{--} \gamma) = 49.5 \%$ and $BR(H^{++} \to U^{++} \gamma) = 49.5 \%$, see Figs. 3 and 4, for the mass of the Higgs boson $m_{\pm \pm}= 500$ GeV, $v_{\chi}=1000$ GeV, and that the particles $U^{\mp \mp}$ decay into $e^{-} P^{-}$ and $e^{+} P^{+}$, whose branching ratios for these particles would be $BR(U^{--} \to e^{-} P^{-}) = 50 \% $ and $BR(U^{++} \to e^{+} P^{+}) = 50 \%$, see Figs. 7 and 8, then  we would have approximately $\simeq 9.1 \times 10^{3}$ events per year for the CLIC, regarding VEV $v_{\chi}=1500$ GeV it will not give any event due to the considerations given above.
Considering now the mass of the Higgs boson $m_{\pm \pm}= 700$ GeV and the  VEV $v_{\chi}=1000$, we then have a total of $1.2 \simeq 10^5$ events per year. Considering now the same signal as above, whose branching  ratios  are equal to $BR(H^{--} \to U^{--} \gamma) = 23.7 \%$ and $BR(H^{++} \to U^{++} \gamma) = 23.7 \%$, see Figs. 3 and 4, for the mass of the Higgs boson $m_{\pm \pm}= 700$ GeV, $v_{\chi}=1000$ GeV, and that the particles $U^{\mp \mp}$ decay into $e^{-} P^{-}$ and $e^{+} P^{+}$, whose branching ratios for these particles would be $BR(U^{--} \to e^{-} P^{-}) = 50 \% $ and $BR(U^{++} \to e^{+} P^{+}) = 50 \%$, see Figs. 9 and 10, then  we would have approximately $\simeq 1.7 \times 10^{3}$ events per year for the CLIC, regarding VEV $v_{\chi}=1500$ GeV  then we will have around $\simeq 5.4  \times 10^{5}$ events per year and   considering the same parameters signals as above, that is, $BR(H^{--} \to U^{--} \gamma) = 47.7 \%$ and $BR(H^{++} \to U^{++} \gamma) = 47.7 \%$, see Figs. 5 and 6, and $BR(U^{--} \to e^{-} P^{-}) = 50 \% $ and $BR(U^{++} \to e^{+} P^{+}) = 50 \%$, see Figs. 11 and 12, then  we would have approximately $\simeq 3.1 \times 10^{4}$ events per year for the CLIC. These results are showed in the Table II.
There are another signals, such as $H^{\mp \mp} \rightarrow U^{\pm \pm} Z  \rightarrow E^{\pm} e^{\pm} e{\pm} e^{\mp}$ and a $H^{\mp \mp} \rightarrow  E^{\pm} e^{\pm}$  these  signals occur with a small probability, the first with a probability of $\simeq 10^{-2}$ and the second with a $\simeq 10^{-5}$ events per year for the following parameters, $v_{\chi}=1500$ GeV, $m_{\pm}=700$ GeV, $\lambda=-0.8$ and the others parameters listed above. As we can see the signals that occur with a large probability are the ones we have considered above. So we have that the high energy electron-positron colliders can be a rich  source for DCHBs. In relation to the signal, $H^{\pm \pm} \to U^{\pm \pm} \gamma$ and $U^{\pm \pm} \to e^{\pm} P^{\pm}$, we conclude  that it is a very striking and important signal, the doubly-charged Higgs boson will deposit six times the ionization energy than the characteristic single-charged particle, that is, if we see this signal we will not only be seeing the DCHBs but also the doubly charged gauge bosons and heavy leptons. It is to notice that there are not standard model backgrounds to this signal. The discovery of a pair of DCHBs will be without any doubt of great importance for the physics beyond the standard model, because of the confirmation of the Higgs triplet representation and indirect verification that there is assymetry in decay rates between matter  and antimatter. Our study indicates the possibility of obtaining a clear signal of these new particles with a satisfactory number of events. 

\end{document}